# Quantum and classical correlations in high temperature dynamics of two coupled large spins

V.E. Zobov*

L. V. Kirensky Institute of Physics, Russian Academy of Sciences, Siberian Branch, 660036, Krasnoyarsk, Russia

## ABSTRACT

We study dynamical correlations of two coupled large spins depending on the time and on the spin quantum numbers. In the high-temperature approximation, we obtain analytical expressions for the mutual informations, quantum and classical parts of correlations. The latter was obtained performing the non-orthogonal projective (POVM) measurements onto the spin coherent states of two spins or one spin, as well as by means of the orthogonal projective measurement of von Neumann. Contribution from quantum correlations is much less than that from classical ones and decreasing with the increase in spin quantum numbers at short times. However, it is not so in a time equal to half of the quantum period.

**Keywords:** dynamical correlations, mutual information, spin coherent states, nuclear magnetic resonance, measurement

## 1. INTRODUCTION

By now, a lot of experimental work demonstrates implementation of quantum algorithms on nuclear spins at high temperatures using the method of nuclear magnetic resonance (NMR)[1]. It is known that entanglement is absent, but the dynamic correlations contain the quantum part[2,3] along with the classic part. To determine the proportion of these parts it is proposed several measures (see for example a review[4]). However, this issue has not been solved yet.

Instead of an arbitrary state, we consider a free induction decay (FID)[5] (a correlation typical for NMR). The FID is the average time evolution of the total X-projection of the spin system, normalized to 1 at the initial time moment:

$$F(t) = \langle \hat{S}_X(t) \rangle / \langle \hat{S}_X(0) \rangle = Tr\{\hat{S}_X \hat{U}(t)\hat{S}_X \hat{U}^{-1}(t)\}/Tr\{(\hat{S}_X)^2\},$$

where $\hat{U}(t) = \exp(-i\hat{H}t/\hbar)$, $\hat{S}_X = \sum_j \hat{S}_{jX}$, $\hat{S}_{j\alpha}$ is α-component of a spin j operator. In the strong static magnetic field (which is typically used in NMR) the nuclear spin polarization is very small[5] at room temperature T: $\beta = \hbar\omega_0/kT \approx 10^{-5} \ll 1$ ($\omega_0$ is the Larmor frequency). Therefore, the density matrix in thermal equilibrium state can be approximated as

$$\hat{\rho}_{eq} = (1 + \beta\hat{S}_Z)/Z,$$

where Z is the partition function. To observe the NMR signal one affects on the system by the radio frequency (RF) pulses, carrying a rotation on angle of $90^0$ about the Y axis of the rotating reference frame (RRF):

$$\hat{\rho}(0) = \hat{Y}\hat{\rho}_{eq}\hat{\bar{Y}} = (1 + \beta\hat{S}_X)/Z. \tag{1}$$

This initial density matrix evolves in time due to the secular dipolar Hamiltonian[5] as

$$\hat{\rho}(t) = \hat{U}(t)\hat{\rho}(0)\hat{U}^{-1}(t) = (1 + \beta\hat{U}(t)\hat{S}_X\hat{U}^{-1}(t))/Z. \tag{2}$$

The polarization of individual spins (initial order) (1) transforms to the correlation between a spin and a local field, created by dipole-dipole interactions:

* rsa@iph.krasn.ru

$$\left[\hat{H}_d, \hat{S}_X\right] = \hbar \sum_{i<j} b_{ij} \left[(3\hat{S}_{iZ}\hat{S}_{jZ} - \vec{\hat{S}}_i \cdot \vec{\hat{S}}_j), \hat{S}_X\right] = i\hbar 3 \sum_{i \neq j} b_{ij} \hat{S}_{iZ} \hat{S}_{jY}. \qquad (3)$$

This correlation has been well studied in NMR. It is responsible for the solid-echo[6], if one irradiates the system by the second $90^0$ RF pulse rotating about the Y axis RRF. It also can be transformed into the energy of the dipole-dipole interactions, if the $45^0$ RF pulse is applied instead[7]. However, these spin-field correlations have not been divided into quantum and classical parts.

Now we study this separation for a model system. We take a system of two large spins $S_1$ and $S_2$ with the dipole-dipole interaction between the projections onto the static magnetic field (axis Z) given by the Hamiltonian:

$$\hat{H}_{SS} = \hbar \frac{J}{S_2} \hat{S}_{1Z} \hat{S}_{2Z}. \qquad (4)$$

To study the limit $S_1, S_2 \to \infty$ we have scaled a constant interaction inversely with $S_2 \geq S_1$, follow the work[8,9]. Such model demonstrates the transition from the case of two spins $S_2 = S_1 = 1/2$, where the dynamic correlations are divided into quantum and classical parts equally[10], to the case of two classical angular momenta, where there are no quantum correlations at all.

## 2. CLASSICAL AND QUANTUM ANGULAR MOMENTA

Before proceed to the analysis of dynamical correlations, we define angular moment states. We characterize the state by the polar $\theta$ and azimuthal $\varphi$ angles on the unit (Bloch) sphere. A classical angular moment $\vec{S}$ has well-defined projections on the coordinate axes:

$$S_z = S \cos\theta, \quad S_y = S \sin\theta \sin\varphi, \quad S_x = S \sin\theta \cos\varphi. \qquad (5)$$

For a quantum angular momentum or a spin we use the basis of the states with a fixed value of the projection on the axis Z:

$$|m\rangle, \qquad (6)$$

where *m* takes 2S+1 values:

$$- S, - S +1, \ldots, S -1, S.$$

It is believed that the most similar to the states of classical angular momenta are the spin coherent states[8,9,11,12] (SCS):

$$|\theta,\varphi\rangle = R(\theta,\varphi)|S\rangle = \sum_{m=-S}^{m=S} \binom{2S}{S+m}^{1/2} (\cos\theta/2)^{S+m} \left(e^{i\varphi} \sin\theta/2\right)^{S-m} |m\rangle, \qquad (7)$$

These states are obtained from the ground state $|S\rangle$ by the rotation operator $R(\theta,\varphi)$, and are a superposition of the states (6) with different *m*. In state (7) the values of the average spin projections:

$$\langle \hat{S}_Z \rangle = \langle \theta,\varphi | \hat{S}_Z | \theta,\varphi \rangle = S \cos\theta,$$

$$\langle \hat{S}_X \rangle = \langle \theta,\varphi | \hat{S}_X | \theta,\varphi \rangle = S \sin\theta \cos\varphi, \qquad (8)$$

$$\langle \hat{S}_Y \rangle = \langle \theta,\varphi | \hat{S}_Y | \theta,\varphi \rangle = S \sin\theta \sin\varphi,$$

have the same values as that of the classical angular momentum (5).

### 2.1 Quantum fluctuations

For arbitrary state we have $\langle(\hat{S})^2\rangle = \langle(\hat{S}_X)^2\rangle + \langle(\hat{S}_Y)^2\rangle + \langle(\hat{S}_Z)^2\rangle = S(S+1)$.

If the state is $|S\rangle$, then $\langle(\hat{S}_Z)^2\rangle = S^2$, therefore $\langle(\hat{S}_X)^2\rangle + \langle(\hat{S}_Y)^2\rangle = S$.

In the derivation of formulas, we shall use the averaging over an isotropic distribution of spin directions:
for the spin

$$\langle(\hat{S}_Z)^2\rangle = \frac{1}{2S+1}\sum_{m=-S}^{m=S} m^2 = S(S+1)/3, \qquad (9)$$

for the classical angular momentum

$$\langle(S_Z)^2\rangle = \frac{1}{2}\int_{-1}^{+1}(S\cos\theta)^2 d\cos\theta = S^2/3. \qquad (10)$$

## 3. MUTUAL INFORMATION OF THE TWO SPINS

Now we return to the system of two large spins $S_1$ and $S_2$, with the spin-spin interaction (4). We use the basis of a direct product $|m_1\rangle_1 \otimes |m_2\rangle_2$. The time dependence of the density matrix (2) can be calculated explicitly as

$$\hat{\rho}(t) = \exp(-i\hat{H}_{SS}t/\hbar)\hat{\rho}(0)\exp(i\hat{H}_{SS}t/\hbar) = \frac{1}{Z}[1 + \frac{\beta}{2}\{\hat{S}_{1+}\exp(-i\tau\hat{S}_{2Z}) + \\ + \hat{S}_{1-}\exp(i\tau\hat{S}_{2Z}) + \hat{S}_{2+}\exp(-i\tau\hat{S}_{1Z}) + \hat{S}_{2-}\exp(i\tau\hat{S}_{1Z})\}] = [1 + \beta\Delta\hat{\rho}(t)]/Z \qquad (11)$$

where $\hat{S}_{j\pm} = \hat{S}_{jX} \pm i\hat{S}_{jY}$ are raising and lowering operators, $\tau = tJ/S_2$ is dimensionless time, $Z = d_1 d_2$, $d_j = 2S_j + 1$.

The mutual information

$$I(\hat{\rho}) = S_H(\hat{\rho}_1) + S_H(\hat{\rho}_2) - S_H(\hat{\rho}), \qquad (12)$$

serves as a measure of correlation between spins. In Eq. (12) $S_H(\hat{\rho}) = -Tr\{\hat{\rho}\log\hat{\rho}\}$ is the von Neumann entropy, $\hat{\rho}_1 = Tr_2\hat{\rho}$, $\hat{\rho}_2 = Tr_1\hat{\rho}$. Reducing the density matrix over states of the one spin, we find the reduced density matrices:

$$\hat{\rho}_1 = \{1 + \beta\hat{S}_{1X}g_2(t)/d_1, \quad \hat{\rho}_2 = \{1 + \beta\hat{S}_{2X}g_1(t)/d_2, \qquad (13)$$

$$g_j(t) = \frac{1}{d_j}\sum_{-S_j}^{S_j}\exp(\pm i\tau m) = \frac{\sin(d_j\tau/2)}{d_j\sin(\tau/2)}. \qquad (14)$$

The von Neumann entropy can be read as

$$S_H(\hat{\rho}) = -Tr\{\hat{\rho}\log\hat{\rho}\} \approx \log Z - \frac{\beta^2}{2\ln 2}Tr(\Delta\hat{\rho})^2, \qquad (15)$$

up to terms of the third order on the small quantity[2,3] β. We find

$$I(\hat{\rho}) \approx S_1(S_1+1)B[1-g_2^2(t)] + S_2(S_2+1)B[1-g_1^2(t)], \qquad (16)$$

where $B = \beta^2/6\ln 2$.

## 4. MUTUAL INFORMATION OF THE TWO CLASSICAL ANGULAR MOMENTA

It is interesting to compare the result (16) with similar result for two classical angular momenta. In this case, instead of the density matrix $\hat{\rho}(t)$ (11) the state of the ensemble is described by a probability distribution of the angles on the Bloch sphere:

$$P_C(\theta_1\varphi_1\theta_2\varphi_2;t) = \frac{1}{(4\pi)^2}[1 + \frac{\beta}{2}\{S_{1+}\exp(-i\tau S_{2Z}) + S_{1-}\exp(i\tau S_{2Z}) + S_{2+}\exp(-i\tau S_{1Z}) + S_{2-}\exp(i\tau S_{1Z})\}], \quad (17)$$

where

$$S_{jZ} = S_j\cos\theta_j, \qquad S_{j\pm} = S_j\sin\theta_j\exp(\pm i\varphi_j).$$

The classical mutual information is

$$J_C(P_C) = S_{Sh}(P_{C1}) + S_{Sh}(P_{C2}) - S_{Sh}(P_C), \qquad (18)$$

Where $S_{Sh}(P_C) = -\iint\iint P_C(\theta_1\varphi_1\theta_2\varphi_2;t)\log P_C(\theta_1\varphi_1\theta_2\varphi_2;t)\sin\theta_2 d\theta_2 d\varphi_2 \sin\theta_1 d\theta_1 d\varphi_1$ is the Shannon entropy. In calculating the reduced distribution, and the Shannon entropy, instead of taking the traces it is necessary to calculate the integrals over the Bloch sphere. The reduced distribution for the first angular momentum (for the second is similar) is

$$P_{C1}(\theta_1\varphi_1;t) = \iint P_C(\theta_1\varphi_1\theta_2\varphi_2;t)\sin\theta_2 d\theta_2 d\varphi_2 = \frac{1}{4\pi}\{1 + \beta S_{1X}g_{C2}(t)\},$$

$$g_{Cj}(t) = \frac{1}{2}\int_{-1}^{+1}\exp(\pm i\tau S_{jZ})d\cos\theta_j = \frac{1}{2}\int_{-1}^{+1}\exp(\pm i\tau S_j\cos\theta_j)d\cos\theta_j = \frac{\sin\tau S_j}{\tau S_j}. \qquad (19)$$

We find the classical mutual information in high-temperature approximation as

$$J_C(P_C) \approx (S_1)^2 B[1-g_{C2}^2(t)] + (S_2)^2 B[1-g_{C1}^2(t)]. \qquad (20)$$

## 5. CLASSICAL CORRELATION OF THE TWO SPINS

The mutual information (16) is a measure of the total correlations, which is the sum of the classical and quantum correlations. To calculate the classical correlation, measurements should be made, or, in other words, we should project a state $\hat{\rho}(t)$ onto some complete system of projectors[4]. We take a system of the SCS (7). These states are considered to be the closest to the state of a classical angular momentum. The completeness properties[11] are

$$\frac{2S+1}{4\pi}\iint|\theta,\varphi\rangle\langle\theta,\varphi|\sin\theta d\theta d\varphi = 1. \qquad (21)$$

Doing POVM (positive - operator - valued - measure) measurement, which is a trace of the product of $\hat{\rho}(t)$ and a projection operator onto the SCS, we find a classical probability distribution function[9,11,12] of the angles:

$$P_{12}(\theta_1\varphi_1\theta_2\varphi_2;t) = d_1 d_2 tr\{|\theta_1,\varphi_1\rangle\langle\theta_1,\varphi_1|\otimes|\theta_2,\varphi_2\rangle\langle\theta_2,\varphi_2|\hat{\rho}(t)\} = \langle\theta_1\varphi_1\theta_2\varphi_2|\hat{\rho}(t)|\theta_1\varphi_1\theta_2\varphi_2\rangle. \qquad (22)$$

Note that the basis of SCS is not orthogonal[11]:

$$\left|\langle\theta,\varphi|\theta',\varphi'\rangle\right|^2 = \cos^{4S}(\Theta/2),$$

$$\cos\Theta = \cos\theta\cos\theta' + \sin\theta\sin\theta'\cos(\varphi-\varphi').$$

The POVM measurements are not equal[4] to the orthogonal projective measurements of von Neumann. Using properties[11] of the SCS:

$$\langle\theta_j,\varphi_j|\hat{S}_{j\pm}|\theta_j,\varphi_j\rangle = S_{j\pm} = S_j\sin\theta_j\exp(\pm i\varphi_j),$$

$$\langle\theta_j,\varphi_j|\exp(\pm i\tau\hat{S}_{jZ})|\theta_j,\varphi_j\rangle = [\cos(\tau/2)\mp i\cos\theta_j(\sin\tau/2)]^{2S_j} \equiv \xi_{j\mp}(t),$$

we find:

$$P_{12}(\theta_1\varphi_1\theta_2\varphi_2;t) = \frac{1}{(4\pi)^2}[1+\frac{\beta}{2}\{S_{1+}\xi_{2+}(t)+S_{1-}\xi_{2-}(t)+S_{2+}\xi_{1+}(t)+S_{2-}\xi_{1-}(t)\}], \quad (23)$$

$$P_1(\theta_1\varphi_1;t) = \oiint P_{12}(\theta_1\varphi_1\theta_2\varphi_2;t)\sin\theta_2 d\theta_2 d\varphi_2 = \frac{1}{4\pi}\{1+\beta S_{1X}g_2(t)\},$$

$$P_2(\theta_2\varphi_2;t) = \oiint P_{12}(\theta_1\varphi_1\theta_2\varphi_2;t)\sin\theta_1 d\theta_1 d\varphi_1 = \frac{1}{4\pi}\{1+\beta S_{2X}g_1(t)\}.$$

Then, follow Eq. (18) we find for the mutual information of the classical probability distribution (23), obtained after the POVM measurements, in the high-temperature approximation:

$$J_{GG}(P_{12}) \approx (S_1)^2 B[f_2(t)-g_2^2(t)] + (S_2)^2 B[f_1(t)-g_1^2(t)], \quad (24)$$

where

$$f_j(t) = \frac{1}{2}\int_{-1}^{+1}\xi_{j+}(t)\xi_{j-}(t)d\cos\theta_j = \sum_{n=0}^{n=2S_j}\binom{2S_j}{n}(\cos^2\tau/2)^{2S_j-n}(\sin^2\tau/2)^n\frac{1}{2n+1}. \quad (25)$$

## 6. QUANTUM PART OF TOTAL CORRELATIONS

If the found classical correlation $J_{GG}(P_{12})$ (24) is now subtracted from the total correlation $I(\hat{\rho})$ (16), the difference gives the quantum part of correlations:

$$Q_{GG} = I(\hat{\rho}) - J_{GG}(P_{12}) \approx (S_1)^2 B[1-f_2(t)] + (S_2)^2 B[1-f_1(t)] + S_1 B[1-g_2^2(t)] + S_2 B[1-g_1^2(t)]. \quad (26)$$

In this case, the measurements were performed symmetrically on both spins. Other measure of classical correlations

$$C_G \approx S_1(S_1+1)B[f_2(t)-g_2^2(t)] + (S_2)^2 B[1-g_1^2(t)] \quad (27)$$

can be obtained, if performed POVM on one spin only (second spin). From which we find the quantum part

$$Q_G = I(\hat{\rho}) - C_G \approx S_1(S_1+1)B[1-f_2(t)] + S_2 B[1-g_1^2(t)]. \quad (28)$$

Note that in systems with a continuous spectrum the Gaussian quantum discord[13,14] was measured, which is used POVM whose elements are the field (boson) coherent states.

## 7. ORTHOGONAL PROJECTIVE MEASUREMENT OF VON NEUMAN

Another measure[4] for classical correlations is proposed using the orthogonal measurement of von Neumann. Let take $S_1 = 1/2$ and the projector

$$\hat{\Pi}_{1\pm} = [1 \pm \hat{\sigma}_{1Z}]/2. \tag{29}$$

The authors of the works[10,15,16] show that it is the optimal choice at $S_1 = S_2 = 1/2$. We found that, at least for short times, this choice is optimal in the case $S_1 = 1/2$, $S_2 \geq 1/2$. Extended to arbitrary times, we find the classical and quantum parts of correlations, respectively:

$$C_O \approx S_2(S_2+1)B\sin^2(\tau/2), \tag{30}$$

$$Q_O = I(\hat{\rho}) - C_O \approx \frac{3}{4}B\left[1 - g_2^2(t)\right]. \tag{31}$$

## 8. DISCUSSION

Figure 1 shows the time dependence of the results (16), (20), and (24) for the mutual informations at different values of the spin quantum number $S_1 = S_2 = S$. The dimensionless time $\tau S = tJ$ and amplitude (in units $2BS^2$) are present. In these units the classical mutual information $J_C(P_C)/(2BS^2)$ (20) is the same for different $S$. In the beginning (at $t = 0$) correlations are absent. As the time growing the correlations scale with the square of the time. At $\tau = tJ/S << 1$ we get:

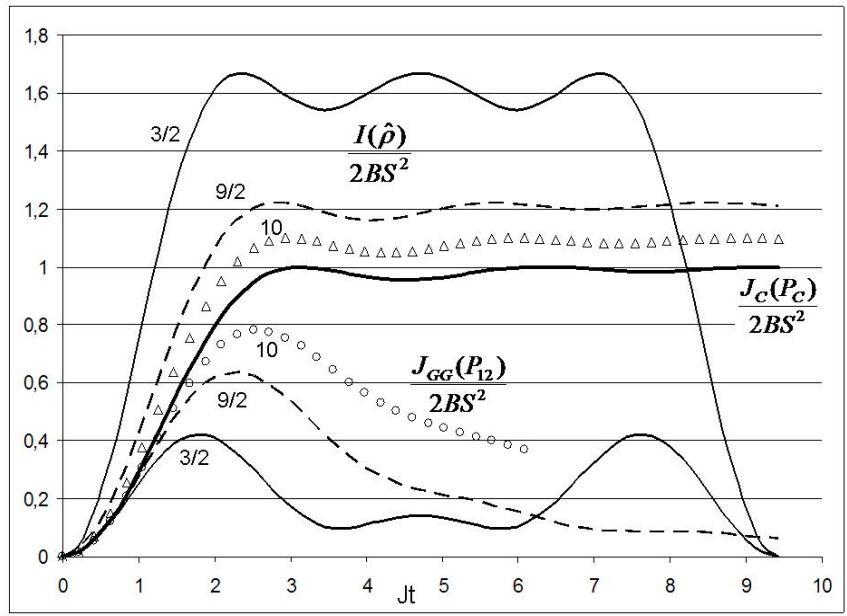

Figure 1. The time evolution of the mutual informations (quantum, classical, and after the POVM measurements onto the SCS) at different values of $S_1 = S_2 = S$ (the number beside the curves).

$$I(\hat{\rho}) \approx \frac{2}{3} B S_1(S_1+1) S_2(S_2+1) \tau^2 = \frac{2}{3} B S_1(S_1+1)(1+\frac{1}{S_2})(Jt)^2, \qquad (32)$$

$$J_C(P_C) \approx \frac{2}{3} B (S_1)^2 (S_2)^2 \tau^2 = \frac{2}{3} B (S_1)^2 (Jt)^2, \qquad (33)$$

$$J_{GG}(P_{12}) = J_C(P_C). \qquad (34)$$

The amplitude of the quantum mutual information (32) exceeds the classical one (33), because due to quantum fluctuations the square of the total momentum is $S(S+1)$ instead $S^2$. At $S \to \infty$ the curves coincide at finite times.

At longer times the classical mutual information $J_C(P_C)$ (20) tends to a limit value. The quantum mutual information $I(\hat{\rho})$ (16) has a different behavior: first, the curve reaches a plateau, but then it decreases to zero at $\tau = 2\pi$. The periodic change in time with the period $T = 2\pi S/J$ is consequence of the discrecity of the energy levels. At $t = T$ differences between the phase shifts of the various energy levels become multiple the $2\pi$, therefore $g^2(T) = 1$ (14) and $I(\hat{\rho}) = 0$. In the case of classical angular momenta the projection of angular momenta on the magnetic field and the energy change continuously, so $g_{Cj}(t) \to 0$ (19) at $t \to \infty$ and the $J_C(P_C)$-curve flattens out. For $S = 3/2$ the periodicity is visible in Fig. 1 ($I(\hat{\rho}) = 0$ at $tJ = 3\pi$). For $S = 9/2$ that happens three times more (at $tJ = 9\pi$). At $S \to \infty$ the period becomes infinite and the finite time curves coincide. Finally, the mutual information $J_{GG}(P_{12})$ (24) of the angles distribution function obtained after POVM, for short times (34) close to the classical $J_C(P_C)$. The long-time curve moves down. It's caused by spreading of the package from the states with different projections $m$ in the SCS (7) (due to a different value of the phase factor $\exp(i\tau \hat{S}_z)$). However, when $t = T$ the package will meet. For $S = 3/2$ this is shown in Fig. 1, for $S = 9/2$ vanishing happens three times more (at $tJ = 9\pi$).

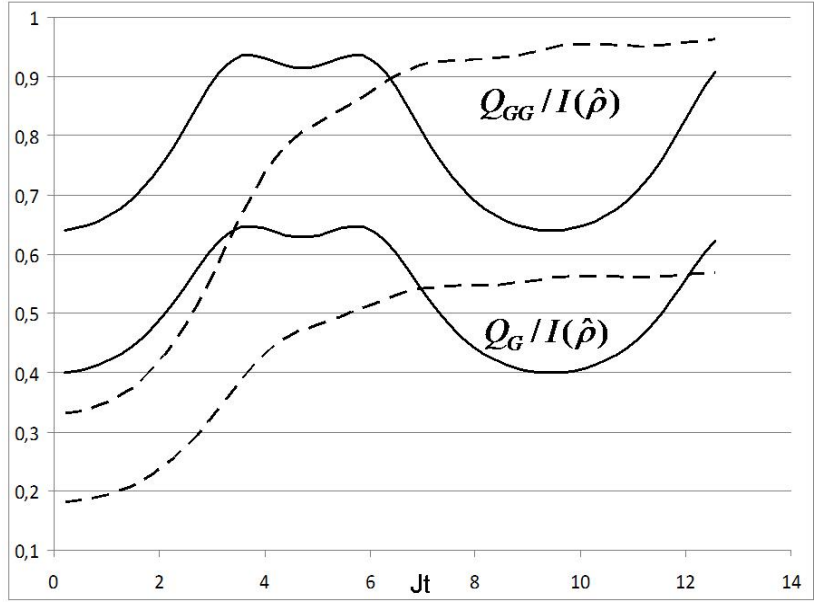

Figure 2. The time evolution of the quantum parts of the correlations under the POVM measurements onto the SCS, as on one or two spins, at two values of $S_1 = S_2 = S$: $S = 3/2$ – solid lines, $S = 9/2$ – dashed lines.

Figure 2 shows the time dependence of the ratios $Q_{GG}/I(\hat{\rho})$ and $Q_G/I(\hat{\rho})$ of the quantum parts to the total correlations at $S_1 = S_2 = S$. From the formulas (16), (26), (28) we find at short time:

$$\frac{Q_{GG}}{I(\hat{\rho})} \approx \frac{2}{S} \text{ and } \frac{Q_G}{I(\hat{\rho})} \approx \frac{1}{S}. \tag{35}$$

These values agree with Fig.2. With increasing time, the proportions of quantum correlations increase and reach maximum values at the time of the order half the period. When $\sin^2(\tau/2) = 1$ only one member remains in the series (25) and we have $f_j(t) = 1/(4S_j + 1)$. Hence, we get the estimate at $S_1 = S_2 = S \gg 1$:

$$\frac{Q_{GG}}{I(\hat{\rho})} \approx 1 - \frac{1}{4S} \text{ and } \frac{Q_G}{I(\hat{\rho})} \approx \frac{1}{2}\left(1 + \frac{3}{4S}\right). \tag{36}$$

We have presented results for the POVM measurements onto the SCS, as one or two spins. The quantum part of the correlations is reduced by about half when measured on one spin. How to measure correctly - not yet been finally resolved the issue. According to Luo[17] as measured on one spin we measure the "semiquantum" correlations (does not get the measure of the classical correlations), which includes part of the quantum correlations.

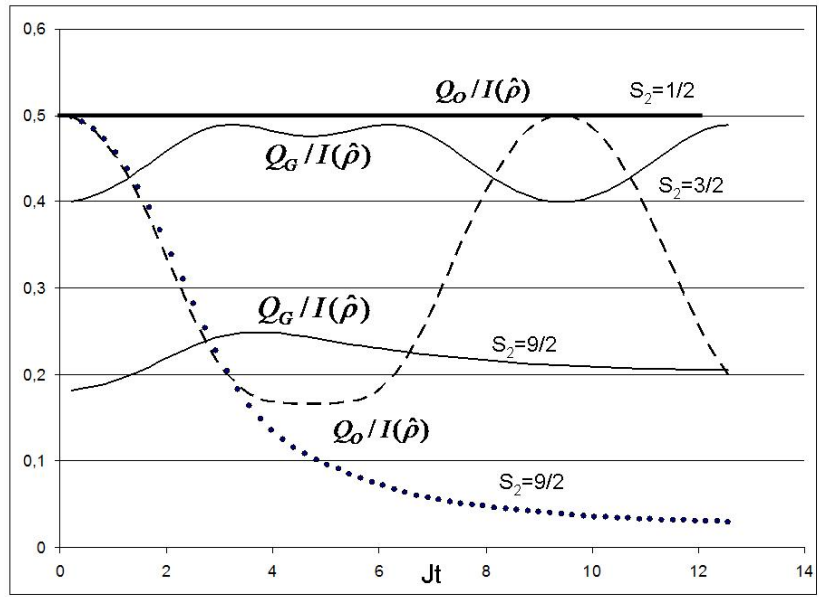

Figure 3. The time evolution of the quantum parts of the correlations at $S_1 = 1/2$ and different values of $S_2$. The results under the POVM measurements on the second spin and (or) under the orthogonal projective measurement on first spin are shown.

Now we compare the result of POVM measurement on the SCS with the result under the orthogonal projective measurement. Figure 3 shows the time evolution of the quantum parts of the correlations at $S_1 = 1/2$. The result of orthogonal measurement at $S_1 = S_2 = 1/2$, as in the work[10], leads to the equality of parts $C_O = Q_O = I(\hat{\rho})/2$ at all times. If $S_2 > 1/2$ there is periodic time dependence. The ratio $Q_O/I(\hat{\rho})$, according (16), (31), decreases with a time and at $\sin^2(\tau/2) = 1$ reaches a minimum value

$$Q_0 / I(\hat{\rho}) \approx 3/(4S_2^2). \quad (37)$$

When POVM performed on the second spin, the time dependence of quantum part $Q_G / I(\hat{\rho})$ at $S_1 = 1/2$ is analogous to the case of large spins. It is seen that if $Jt < 2$ this quantum part of the correlations has a lower value than this part from the orthogonal measurement, whereas at $t \sim T/2$ - on the contrary.

Let us examine why this is. At short time the density matrix (11) has the form:

$$\hat{\rho}(t) \sim \frac{1}{Z}[1 + \beta(\hat{S}_{1X} + \hat{S}_{2X}) + \frac{\beta tJ}{S_2}\{\hat{S}_{1Y}\hat{S}_{2Z} + \hat{S}_{2Y}\hat{S}_{1Z}\}]. \quad (38)$$

After orthogonal projection (29) we lose the summand $\hat{S}_{1Y}\hat{S}_{2Z}$ and half correlations for any value of the spins. When POVM performed on the second (big) spin, then the both terms contribute. Losing only quantum fluctuations when calculating the average of the square of the spin operator components: $S_2^2/3$ instead $S_2(S_2+1)/3$. In the limit of large spin $S \to \infty$ the fluctuations and the differences disappear.

At long times after the orthogonal projection we lose in (11) the same term (as in (38)), but his contribution changes in size (at $S_1 = 1/2$, $S_2 \geq 1/2$). The contribution "the small field from the small spin turns the big spin" remains, whereas the contribution "the big spin turns the small" loses.

At $t = T$ the correlations reduce to zero in the quantum system. At $t = T/2$ they reach their maximum values, as shown in the graphs of mutual informations in Fig.1 and Fig,2. What is the proportion of quantum and classical correlations depends on the method of measurement. Look at the density matrix (11) at $t = T/2$ ($\tau = \pi$):

$$\hat{\rho}(T/2) == \frac{1}{Z}[1 + \frac{\beta}{2}\{\hat{S}_{1+}\exp(-i\pi\hat{S}_{2Z}) + \hat{S}_{1-}\exp(i\pi\hat{S}_{2Z}) + \hat{S}_{2+}\exp(-i\pi\hat{S}_{1Z}) + \hat{S}_{2-}\exp(i\pi\hat{S}_{1Z})\}]. \quad (39)$$

When the value of the spin projection changes on the unit the function $\exp(i\pi S_{jZ})$ changes its sign. The terms are as follows:

for integer spins we have

$$\beta\{\hat{S}_{1X}(-1)^{m_2} + \hat{S}_{2X}(-1)^{m_1}\};$$

for half-integer spins -

$$\beta\{\hat{S}_{1Y}(-1)^{m_2} + \hat{S}_{2Y}(-1)^{m_1}\}.$$

Under orthogonal measuring with the projection onto the basis (6) on one spin (the first) one term will contribute, and we get to the parts of classical and quantum correlations, respectively:

$$\frac{C_O}{I(\hat{\rho})} = \frac{S_2(S_2+1)}{S_1(S_1+1)+S_2(S_2+1)} \text{ and } \frac{Q_O}{I(\hat{\rho})} = \frac{S_1(S_1+1)}{S_1(S_1+1)+S_2(S_2+1)}. \quad (40)$$

When POVM measured on the SCS, situation is more complicated. Such state (7) is formed by sum of states with different projections onto the axis Z, which will give an alternating series, summing at

$$\xi_{j\mp}(T/2) = [\cos\theta_j]^{2S_j}.$$

As a result, after POVM measuring we obtain lower values:

$$\frac{C_{G2}}{I(\hat{\rho})} = \frac{S_1(S_1+1)/(4S_2+1)+S_2^2}{S_1(S_1+1)+S_2(S_2+1)} \text{ and } \frac{C_{GG}}{I(\hat{\rho})} = \frac{S_1^2/(4S_2+1)+S_2^2/(4S_1+1)}{S_1(S_1+1)+S_2(S_2+1)}. \quad (41)$$

We get the estimates (36) for the proportions of the quantum correlations at $S_1 = S_2 = S \gg 1$.

## 9. CONCLUSION

The use of high-temperature approximation allows to obtain analytical formulas to describe the time dependence of classical and quantum correlations between the two spins with arbitrary quantum numbers.

Performing the POVM measurement onto the spin coherent states we shown decreasing at $t << T$ in the proportion of quantum correlations with increasing spin quantum number S, i.e. under the transition to the classical angular momenta.

We found that for spins with finite quantum numbers the quantum part of correlations increases with time and may outperform the classical part of correlations. The presence of quantum correlations will allow realizing quantum computing on large spins (qudits).